\documentclass[12pt]{article}
\usepackage{graphicx}

\newcommand\pubnumber{xxxx-xxx-xx}
\newcommand\pubdate{\today}

\def\icrr{Kamioka Observatory, Institute for Cosmic Ray Research,
    the University of Tokyo, Higashi-Mozumi, Kamioka, Hida, Gifu, 506-1205, Japan}
\def\ipmu{ Kavli Institute for the Physics and Mathematics
    of the Universe (WPI), the University of Tokyo, Kashiwa,
    Chiba, 277-8582, Japan}

\def\Title#1{\begin{center} {\Large #1 } \end{center}}
\def\Author#1{\begin{center}{ \sc #1} \end{center}}
\def\Address#1{\begin{center}{ \it #1} \end{center}}

\newcommand\pubblock{\rightline{\begin{tabular}{l} \pubnumber\\
         \pubdate  \end{tabular}}}
\newenvironment{Abstract}{\begin{quotation}  }{\end{quotation}}
\newenvironment{Presented}{\begin{quotation} \begin{center} 
             PRESENTED AT\end{center}\bigskip 
      \begin{center}\begin{large}}{\end{large}\end{center} \end{quotation}}
\def\Acknowledgements{\bigskip  \bigskip \begin{center} \begin{large}
             \bf ACKNOWLEDGEMENTS \end{large}\end{center}}
\def\beq{\begin{equation}}
\def\eeq#1{\label{#1}\end{equation}}
\def\eeqn{\end{equation}}

\def\beqa{\begin{eqnarray}}
\def\eeqa#1{\label{#1}\end{eqnarray}}
\def\eeqan{\end{eqnarray}}

\let\bar=\overbar

\def\Dslash{\not{\hbox{\kern-4pt $D$}}}
\def\dslash{\not{\hbox{\kern-2pt $\del$}}}

\def\msb{{\bar{\ssstyle M \kern -1pt S}}}

\begin{document}
\begin{titlepage}
\pubblock

\vfill
\Title{Direct Dark Matter Search with XMASS-I}
\vfill
\Author{ Masaki Yamashita for the XMASS collaboration}
\Address{\icrr}
\Address{\ipmu}
\vfill
\begin{Abstract}
 XMASS-I uses single phase liquid xenon  technology for aiming at the direct detection of dark matter.  The detector observes only scintillation light  by 2 inch 642 PMTs which are placed in sphere shape around  an active volume.  With its large mass  target and high photoelectron yield, we conducted a search for dark matter by annual modulation with 832 kg $\times$ 359.2 days  exposure of data.   We find no modulation signal in the data  so that we set an upper limit 4.3$\times10^{-41} \rm{cm}^{2} $ at WIMP mass of 8 GeV/$c^{2}$ which excluded an interpreted  DAMA/LIBRA allowed region.
 \end{Abstract}
\vfill
\begin{Presented}
The 12$^{th}$ Conference on the Intersections of Particle and Nuclear Physics in Vail, CO, U.S.A. \\
(CIPANP2015), May 19-24, 2015 

\end{Presented}
\vfill
\end{titlepage}
\def\thefootnote{\fnsymbol{footnote}}
\setcounter{footnote}{0}
\section{Introduction}
The XMASS project  is a multi-purpose experiment for the underground physics and its targets are dark matter search, low energy solar neutrinos and neutrinoless double beta decay with ton scale fiducial volume of liquid xenon (LXe) \cite{2000:suzuki}.  
The current phase of XMASS is using 832 kg liquid xenon detector mainly for the dark matter search prior to the second stage of 5 ton detector with 1ton fiducial volume  (XMASS 1.5).  Based on the results and experiences which will be obtained by those detectors, the final  detector will be employed for the study of low energy solar neutrinos and the further investigation of dark matter. 

Recently, XMASS carried out searches for dark matter as well as Axion  by taking advantages of  its high photoelectron (PE) yield ($\sim$15 PE/keV) and low background \cite{XMASS1, XMASS3, XMASS_SW, XMASS4}.
The detector looks for not only nuclear recoil from WIMP but also electrons via interactions with other candidates such as Axion like particles and Super-WIMPs. 
 In addition to those searches,  we have also explored dark matter  by annual modulation in both model independent and dependent way \cite{XMASS_MOD}.  In this paper, we report the model dependent annual modulation search, namely  for WIMP via elastic scattering via nuclei, with high exposure data  which is another feature of a large mass detector. 
DAMA/LIBRA observed a positive evidence of the annual modulation in the signal count rate and its statistical significance exceeds 9 $\sigma$ in their 1.33 ton$\cdot$year of data taken over the 14 annual cycles with the 100 kg to 250 kg of NaI(Tl) scintillator crystals \cite{dama}.    As the rotation of the Earth around the Sun produces a modulation in the signal count rate of dark matter, it can be observed by the terrestrial detector \cite{Drukier}.  
The liquid noble gas detector has an advantage in its large mass and XENON100 recently conducted annual modulation search with two phase Xe detector in their 34 kg $\times$ 224.6 days data \cite{XENON_MOD}.
The XMASS detector is able to catch up with the DAMA/LIBRA exposure in less than two years using 832 kg of liquid xenon (LXe) as a target material. It is particularly important  because this is the first time exploration with a compatible exposure in the different environment and underground site.  Moreover comparing to their energy threshold 2 keV$_{\rm{ee}}$, the lower energy threshold of about 1 keV$_{\rm{ee}}$ achieved by XMASS. This translates directly into  higher sensitivity for dark matter searches.
\section{XMASS experiment}
 The XMASS detector is a large LXe scintillation detector and located at the Kamioka Observatory (overburden 2700 m water equivalent) in Japan. The detailed detector design, calibration system and performances are described in \cite{XMASS_Det}. A water tank, 10 m in diameter and 10.5 m height, which is equipped with 72 Hamamatsu H3600 20-inch photomultiplier tubes (PMTs)  acts as both an active muon veto and passive radiation shield against neutrons and gamma rays from the rock.  The LXe is contained in a vacuum insulated Oxygen Free High Conductive (OFHC) copper  vessel. The total amount of LXe in the active volume is 832 kg and the total mass of liquid in the detector is 1050 kg. Hexagonal 642 Hamamatsu R10789 PMTs  are mounted in an approximate sphere OFHC holder with a radius of about 40 cm. The PMT's photo-cathode highly cover the inner surface of the detector with 62.4$\%$. 
 The gains of PMTs is continuously monitored by a blue LED embedded in the inner surface of the detector. Energy calibrations are carried out by regularly inserting a $^{57}$Co micro source \cite{XMASS_Cal} into the working detector and occasionally also $^{55}$Fe, $^{109}$Cd, $^{241}$Am sources.

\section{Data taking and event selection}
   The data accumulated between November-2013 and March-2015 were used for this analysis . Temperature and pressure of LXe were stable in the range of  172.6-173.0 $K$ and  0.162 - 0.164 MPa absolute for the data.  After removing periods of operation with excessive PMT noise, unstable power supply on data acquisition system, or abnormal trigger rate, the total live time becomes 359.2 days.

  In this paper,  three different energy scales were used: 1) keV$_{^{57}\rm{Co}}$ denotes an energy scale obtained by dividing the observed photoelectron (PE) by the PE/keV at 122 keV. Therefore keV$_{^{57}\rm{Co}}$ is proportional to total PE. 2) keV$_{\rm{ee}}$ represents, an electron equivalent energy incorporating all the gamma-ray calibrations and 3) keV$_{\rm nr}$ denotes the nuclear recoil energy using a factor converting  the energy scale from keV$_{^{57}\rm{Co}}$ to keV$_{\rm nr}$ which was taken from \cite{Leff}. The 'PE-scaled' energy keV$_{^{57}\rm{Co}}$ was used throughout the analysis and the final results were translated into keV$_{\rm{ee}}$ and keV$_{\rm nr}$. The energy threshold for this analysis is 0.5 keV$_{^{57}\rm{Co}}$ ($\sim$8 PE) which corresponds to 1.1 keV$_{\rm{ee}}$ and 4.8 keV$_{\rm nr}$. 

\begin{figure}[htbp]
\centering
\includegraphics[width=0.45\textwidth]{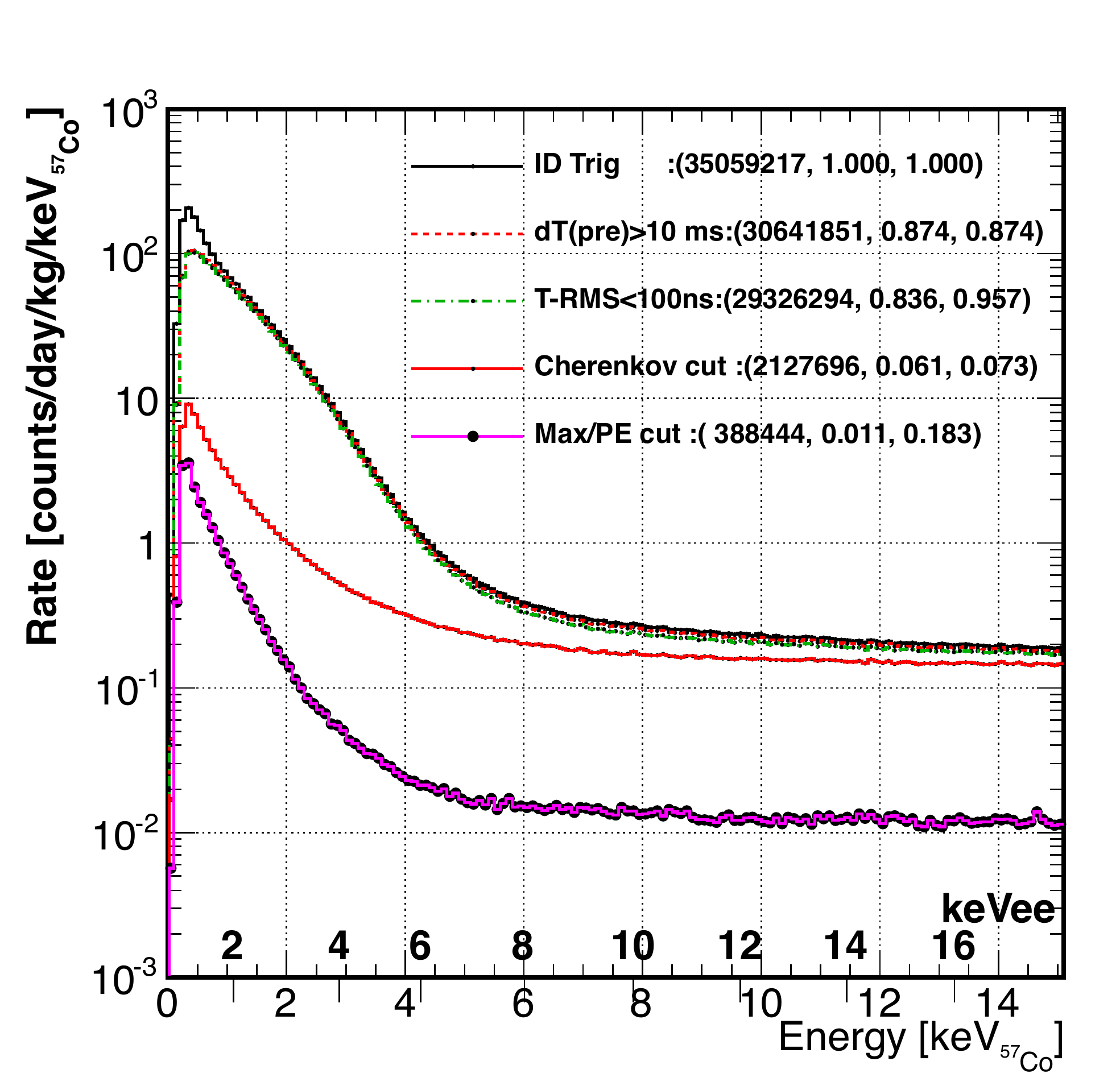}
\includegraphics[width=0.45\textwidth]{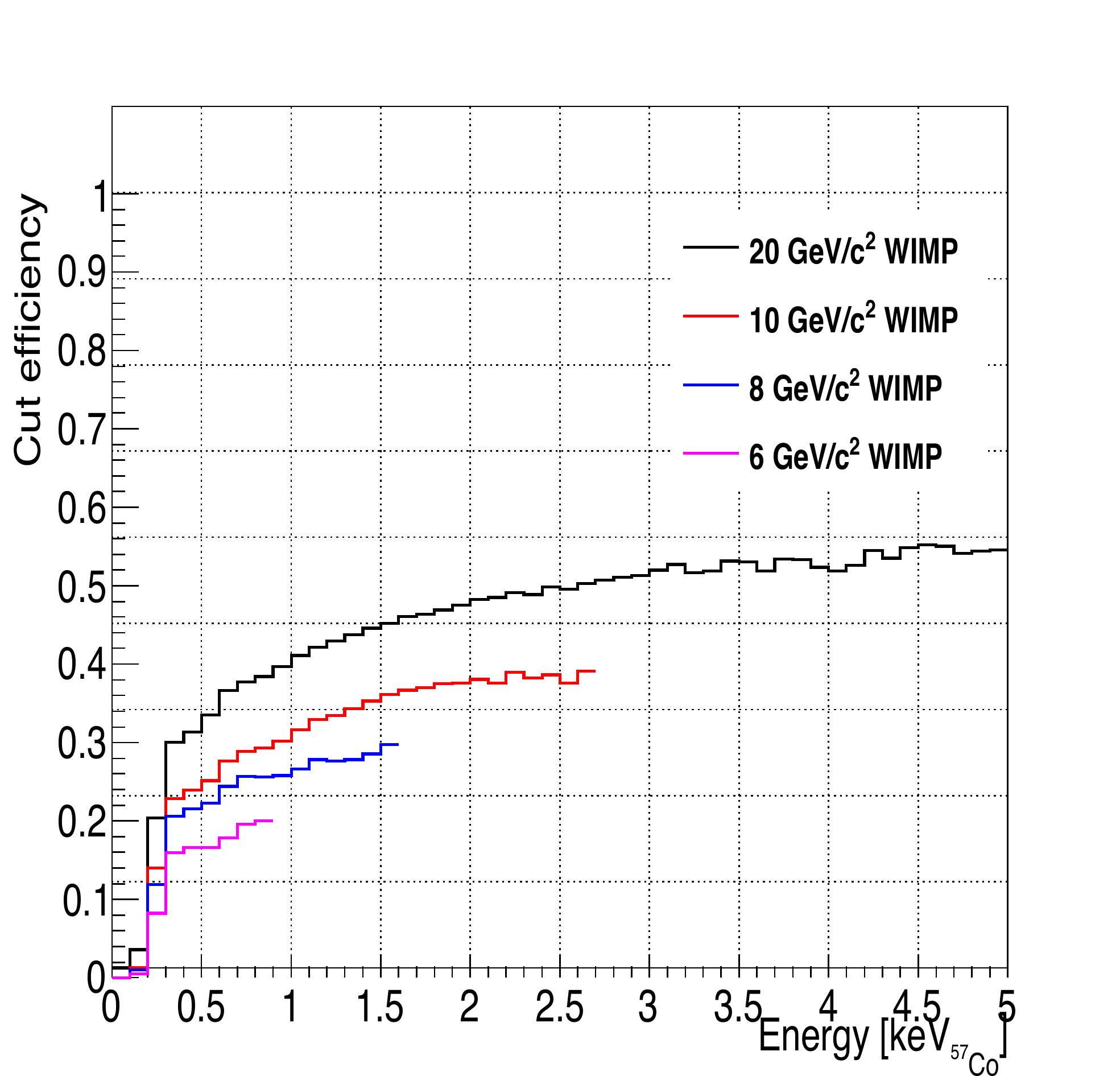}
\caption{Energy spectrum after each event selection for total exposure (left). Total cut efficiency for different WIMP masses from 6 to 20 GeV/$c^{2}$ (right).
These curves end at some point due to their small statistics. } 
\label{fig:histo}
\end{figure}

 Simple data selections were applied for the data analysis and they are described in \cite{XMASS1} except 'Max/Total PE' cut. 
 Figure~\ref{fig:histo} (left) shows the energy spectrum after each data selection.  'ID Trig'  requires  events with 4 PMT hits in a 200 ns coincidence timing window without a muon veto. In order to remove events caused by after pulses of bright events, for example,  induced by high energy gamma-rays or alpha particles, events occurring within 10 ms from the previous event (dT(pre)) or having a variance in their hit timings of greater than 100 ns are removed(T-RMS).  A 'Cherenkov cut' removes events which produce light predominantly through Cherenkov emission, in particular the beta decays of $^{40}$K in the PMT photocathode. Events for which more than 60\,\% of their PMT hits arrive in the first 20\,ns were classified as Cherenkov-like events  \cite{XMASS1}.  Finally, a 'Max/Total PE' cut removes the events that occurred in front of PMTs near the inner surface of the detector. Events with a large ratio of the PE in the PMT with the largest PE count to the total PE observed by all PMTs in the event were removed.  
 After applying Cherenkov cut, this last cut, as a function of PE, was designed to keep 50\% of an uniformly distributed 20 GeV WIMP MC signal, therefore the efficiencies are different for other WIMP masses.
 The cut value varied about 0.2  at 8 PE and about 0.07 at 50 PE and the total efficiencies for each WIMP masse are shown in Fig.~\ref{fig:histo} (right).
 \begin{figure}[htbp]
\centering
\includegraphics[width=0.45\textwidth]{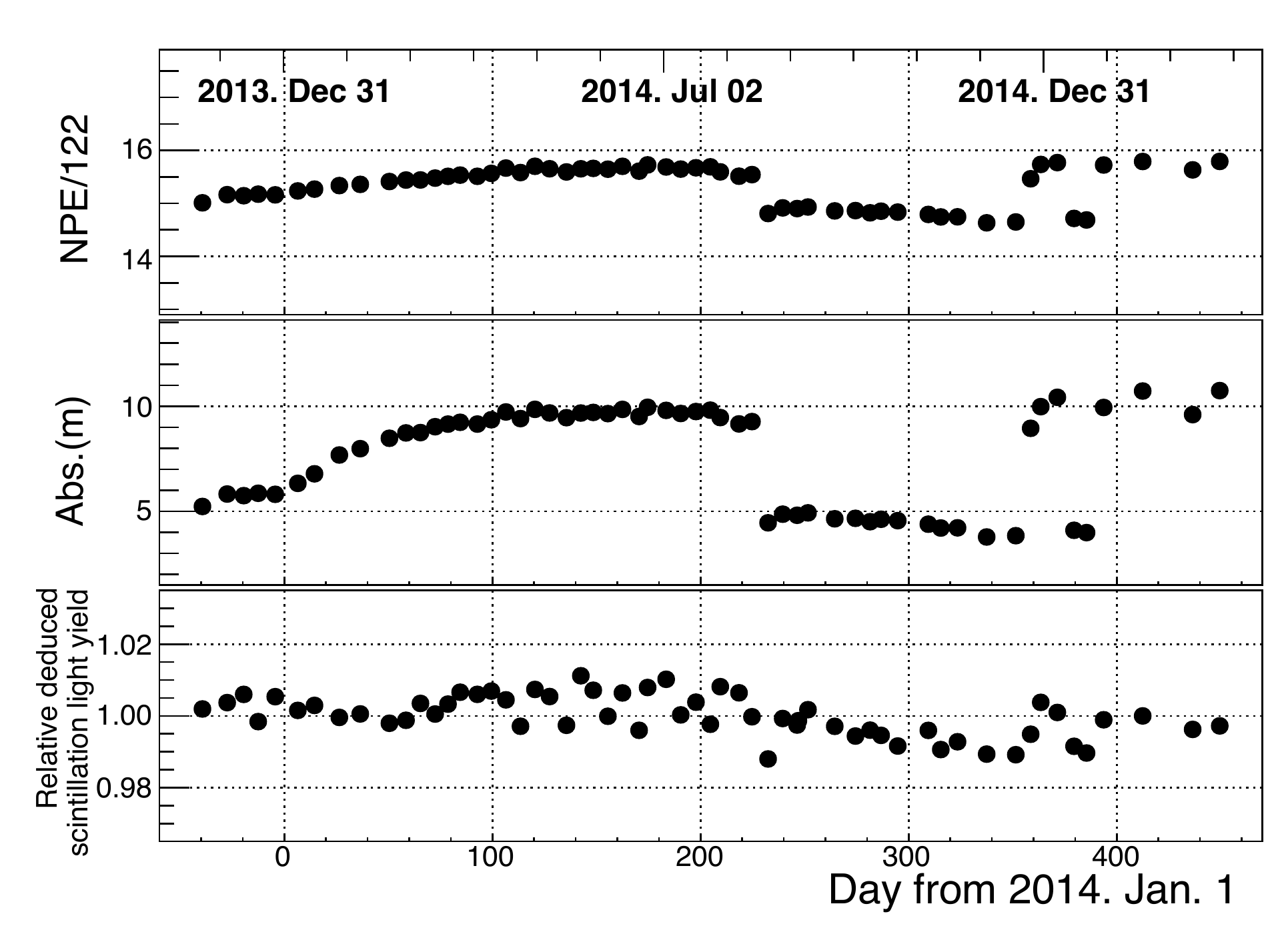}
\includegraphics[width=0.45\textwidth]{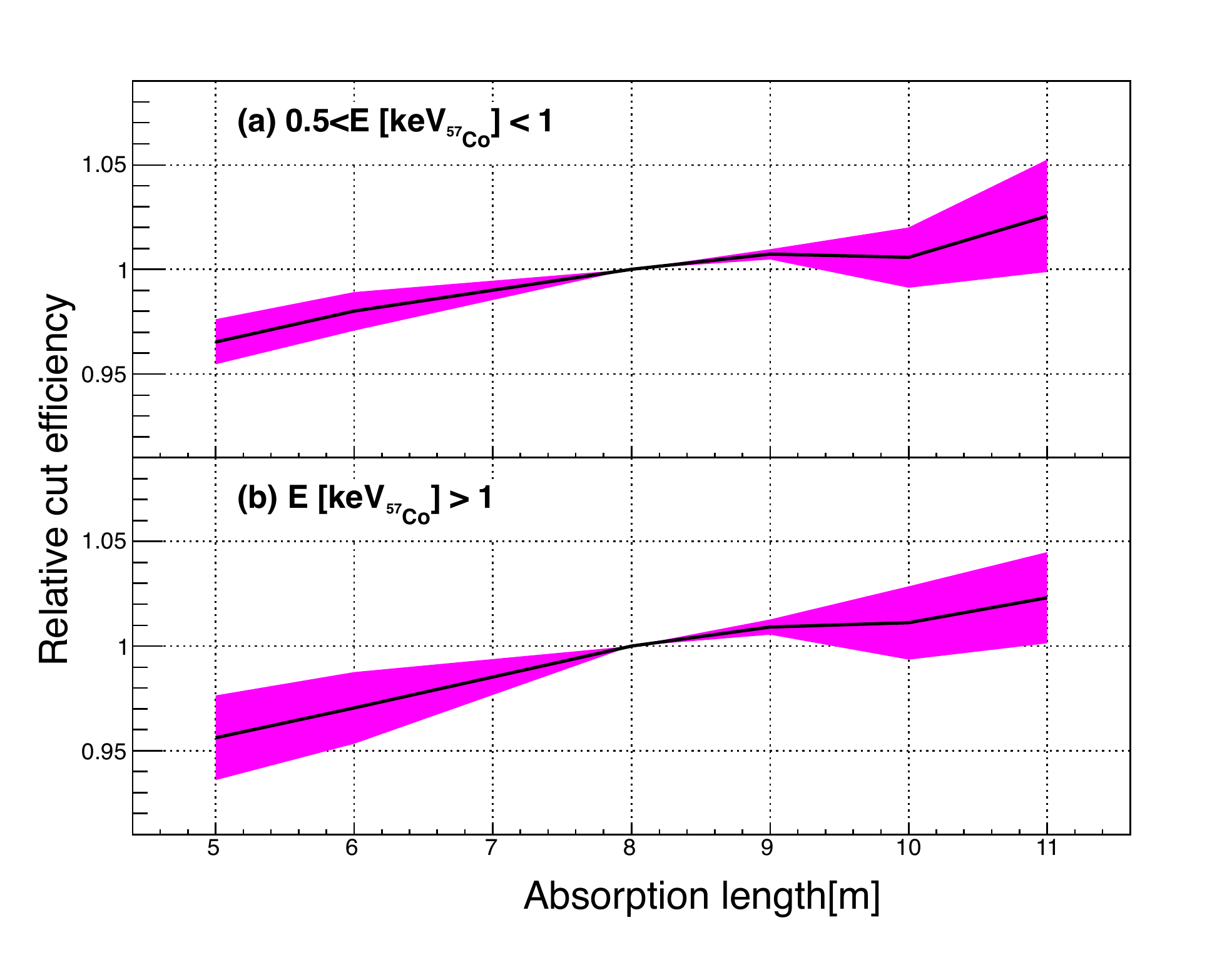}
\caption{PE yield monitoring by inserting $^{57}$Co 122 keV gamma ray source. Optical parameters such as absorption length and deduced intrinsic scintillation light yield were obtained by comparing calibration data with Monte Carlo (MC) simulation. } 
\label{fig:stability}
\end{figure}

 As described in \cite{XMASS_MOD},  the $^{57}$Co calibration data were monitored  to track PE yield and optical properties of the LXe and they are shown in  Fig.~\ref{fig:stability} (left).  
 The central panel in Fig.~\ref{fig:stability} (left) traces the absolute absorption length from about 4 m to 11 m and its bottom panel shows the relative change in the intrinsic light yield with its variation staying within $\pm 0.6\%$ over the entire data taking period. 

 The time dependence of the PE yield affects the efficiency of the cuts. Therefore we evaluate the absorption length dependence of the relative cut efficiencies through MC simulation as shown in  Fig.~\ref{fig:stability}.  The relative cut efficiencies were evaluated for two different energy range to take into account the energy dependency.  We normalize the overall efficiency at an absorption length of 8 m for different energy range with correlated uncertainty due to the position dependence which are described as width of these bands and this is the dominant systematic uncertainty in this analysis. The second largest contribution comes from a gain instability of the FADCs (CAEN V1751)  between April 2014 and September 2014 due to a different initialization method used in that period. It contributes 0.3\% uncertainty to energy scale. Other effects from LED calibration, trigger threshold stability, timing calibration were negligible. 
 
 To retrieve the annual modulation amplitude from the data, the least squares method for the time-binned data
was used. The dataset was divided into 40 time bins ($t_{bins}$) with roughly 10 days of live time each,
and each of these time bins were then further divided into energy bins  ($E_{bins}$) with a width of 0.1 keV$_{^{57}\rm{Co}}$. Two independent fits were performed, differing in their treatment of the systematic uncertainties. Both of them fit all the energy and time bins simultaneously.
 Method 1 introduced a 'pull term' $\alpha$,  and its $\chi^{2}$ was defined as:
\begin{equation}
\chi^2 = \sum\limits_{i}\limits^{E_{bins}} \sum\limits_{j}\limits^{t_{bins}} 
\frac{(\rm{R}^{data}_{i,j}- \rm{R}^{ex}_{i,j} - \alpha K_{i,j})^2}
{\sigma(\rm{stat})^2_{j}}+\alpha^{2} (\rm{method 1})
\end{equation}
where $R_{i,j}^{\rm data}$, $R_{i,j}^{\rm ex}$ and $\sigma(\rm{stat)}$ are the data, expected
event rate and the statistical error in the $i$-th  energy and $j$-th time bin which is relative to January 1, 2014, respectively. $K_{i,j}$ represents
the 1$\sigma$ systematic error on the expected event rate in that bin. Method 2 introduces a covariance matrix to propagate the effects of the systematic error and its $\chi^{2}$ was defined as
\begin{equation}
\chi^2 =\sum_{k,l}^{N_{\rm bins}} (R_{k}^{\rm data}-R_{k}^{\rm ex}) 
(V_{\rm stat}+V_{\rm sys})^{-1}_{kl} 
(R_{l}^{\rm data}-R_{l}^{\rm ex}) \\
 (\rm{method 2}) ,
\end{equation}
where $N_{\rm bins} (= E_{bins} \times t_{bins})$ is the total number of  bins,
$R_{k}^{\rm data(ex)}$ are the data (expected) event rate  $\rm{R}^{data(ex)}_{i,j}$  for the $k = i E_{bins}+j$.
 The matrix $V_{\rm stat}$  contains the statistical uncertainties of the bins, and $V_{\rm sys}$ is the covariance Matrix of the systematic uncertainties as derived from MC simulation. 
As the WIMP mass $m_x$ determines the recoil energy spectrum, the expected modulation amplitudes $A_i$ of the energy bins become a function of the WIMP mass: $A_i(m_x)$.  The expected rate in a bin then becomes:
\begin{equation}
R_{i,j}^{\rm ex}= C_{i} + \sigma_{\chi n} \times A_{i}(m_{\chi}) \cos 2\pi (t_j-t_{0})/T \ ,
\label{eq:MD}
\end{equation}
where  $\sigma_{\chi n}$ is  the WIMP-nucleon cross section.  In these fitting procedure the modulation period $T$ was fixed to one year and phase $t_{0}$ to the expected value of  and 152.5 days ($\sim$2nd of June), the time for maximizing the Earth velocity relative to the WIMP distribution.
\section{Results}
\begin{figure}[htbp]
\centering
\includegraphics[width=0.7\textwidth]{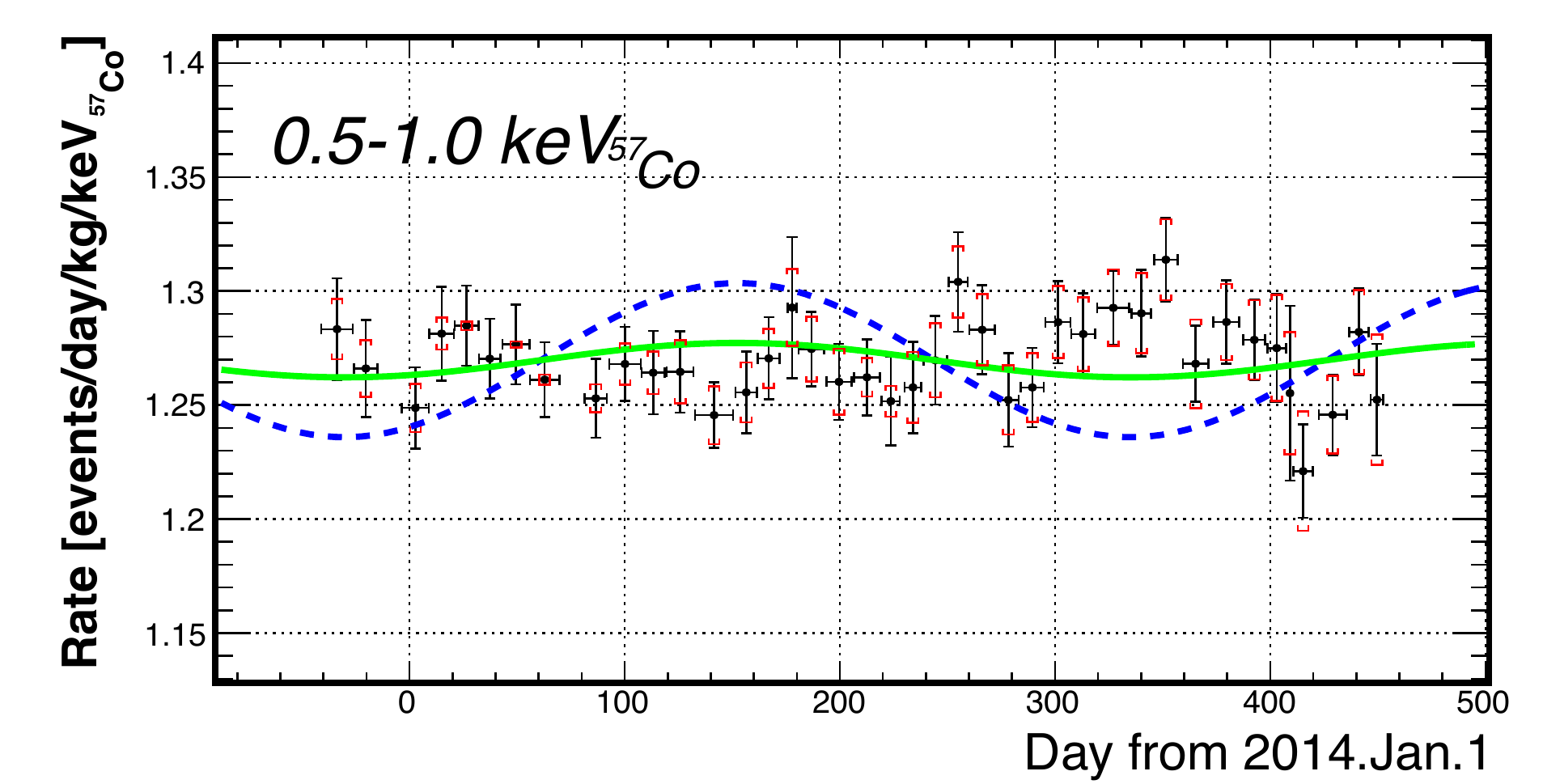}
\caption{Observed count rate as a function of time in the 0.5 - 1.0 keV$_{^{57}\rm{Co}}$ (4.8-8.0 keVnr) energy bin. Error bars represent statistical errors only for each period while square brackets indicate the 1 $\sigma$ systematic error caused by the absorption length difference during the run.
The solid and dashed  curves indicate the expected count rates assuming 7 and 8 GeV/$c^{2}$ WIMPs  with WIMP-nucleon cross section of 2$\times10^{-40} \rm{cm}^{2}$. Detector response and efficiency are accounted in the calculation.}
\label{fig:rate}
\end{figure}
To obtain the WIMP-nucleon cross section the data was fitted in the energy range of  0.5-15 keV$_{^{57}\rm{Co}}$.  We assume a standard spherical isothermal galactic halo model with the most probable speed of $v_{0}$=220~km/s, the Earth's velocity relative to the dark matter distribution of $v_{E} = 232 + 15 ~{\rm sin} 2\pi(t - t_{0})/T$~km/s, and a galactic escape velocity of $v_{esc}$ = 650 km/s, a local dark matter density of 0.3 GeV/cm$^{3}$, following \cite{Lewin}. 

The observed count rate after cuts as a function of time in the energy region of 0.5-1.0 keV$_{^{57}\rm{Co}}$ is shown in Fig.~\ref{fig:rate}.  The size of systematic error caused by the relative cut efficiencies are also shown. The  expected count rate for WIMP masses of 7 and 8 GeV/$c^{2}$ with a cross section of 2$\times10^{-40}$ cm$^{2}$  for the spin independent case  are also shown as a function of time after all cuts. 
The systematic error of the efficiencies comes from the uncertainty of LXe scintillation decay time of 25$\pm$1 ns \cite{XMASS1} and  is estimated as about 5\% in this analysis.

\begin{figure}[htbp]
\centering
\includegraphics[width=0.7\textwidth]{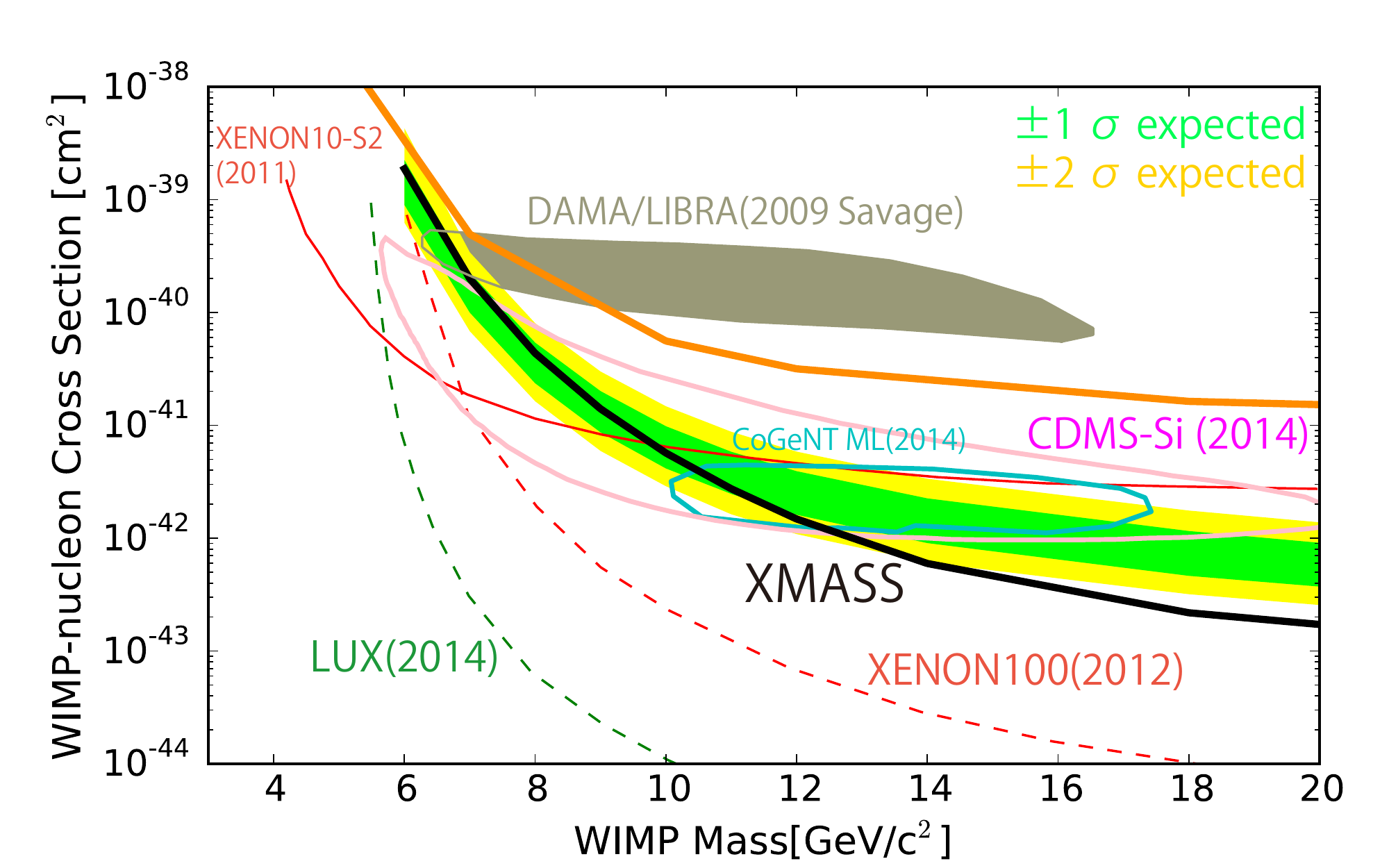}
\caption{ Limits on the spin-independent elastic WIMP-nucleon cross section as a function of WIMP mass. The black line shows the XMASS 90 \% C.L. exclusion from the model dependent  annual modulation analysis. The green and yellow bands represent the $1 \sigma$ and $2 \sigma$ expected 90\% exclusion caused by the statistical and systematic error from dummy samples. Limits  and favored regions from other searches are also shown \cite{XMASS1, XMASS_MOD, LUX, XENON100,SAVAGE, CoGeNT,CDMS-Si,Xe10}. }
\label{fig:MD}
\end{figure}

 As both methods found no significant signal, the 90\% C.L. upper limit by method 1 on the WIMP-nucleon cross section is shown in Fig.~\ref{fig:MD}. The $-1\sigma$ scintillation efficiency of \cite{Leff} was used  to obtain a conservative limit.
 To evaluate the sensitivity, 10,000 dummy samples with the same statistical and systematic errors as data but without modulation were prepared. The $\pm1 \sigma$ and  $ \pm2\sigma$ bands in Fig.~\ref{fig:MD} outline the expected 90\% C.L. upper limit band for the no-modulation hypothesis using these dummy samples.  
The result excludes the DAMA/LIBRA allowed region as interpreted  in  \cite{SAVAGE} for WIMP masses higher than 8 GeV/$c^{2}$. This limit is robust against the difference of analysis methods (less than 10\% for the cross section)  and astrophysical assumptions (upper limit of 5.4$\times10^{-41} \rm{cm}^{2}$ in the case of  $v_{esc}$ = 544~km/s \cite{RAVE}).
 The best fit parameters in a wider mass range is a cross section of 3.2$\times10^{-42}$ $\rm{cm}^{2}$ for a WIMP mass of 140 GeV/$c^{2}$. This yields a  statistical significance of 2.7$\sigma$, however, in this case, the expected unmodulated event rate  exceeds the total observed event rate by a factor of 2, therefore these parameters were deemed unphysical.   
 
 \section{Summary and Conclusion}
 XMASS carried out the annual modulation analysis with large exposure of 832kg $\times$ 359.2 days with  high PE yield LXe detector. 
 For the model dependent case,  the exclusion upper limit 4.3$\times10^{-41} \rm{cm}^{2} $ at 8 GeV/$c^{2}$ was obtained and the result excluded DAMA/LIBRA allowed region in the  higher than that WIMP mass.

\Acknowledgements
We gratefully acknowledge the cooperation of Kamioka Mining
and Smelting Company. 
This work was supported by the Japanese Ministry of Education,
Culture, Sports, Science and Technology, Grant-in-Aid
for Scientific Research, 
JSPS KAKENHI Grant Number, 19GS0204, 26104004, and partially
by the National Research Foundation of Korea Grant funded
by the Korean Government (NRF-2011-220-C00006).

\end{document}